\documentclass[doublespacing]{elsart}
\usepackage{epsfig}

\usepackage{amssymb}

\begin{document}

\begin{frontmatter}

\title{Mechanical Lamb-shift analogue for the Cooper-pair box}

\author[label1]{A. D. Armour,}
\author[label2]{M. P. Blencowe,\corauthref{cor}}
\author[label3]{K. C. Schwab}
\address[label1]{Blackett Laboratory, Imperial College, London SW7 2BW, U.K.}
\address[label2]{Department of Physics and Astronomy, Dartmouth College,
Hanover, New Hampshire 03755}
\address[label3]{Laboratory for Physical Sciences, College Park, 
Maryland 20740}
\corauth[cor]{Corresponding author:  Fax: +(603)-646-1446,
email: m.blencowe@dartmouth.edu}

\begin{abstract}
 We estimate the correction to the Cooper-pair box energy level splitting 
 due to the quantum motion of a 
   coupled micromechanical gate electrode. While the 
   correction due to zero-point motion is very small, it should be 
   possible to observe thermal motion-induced corrections to the 
   photon-assisted tunneling current.         
    
\end{abstract}

\end{frontmatter}

In a recent experiment \cite{nakamura}, 
the coherent control of macroscopic quantum superposition states in a
Cooper box was demonstrated. This represents an important advance 
towards the realization of a solid state quantum computer 
\cite{makhlin}. The ability to manipulate the Cooper box states also 
allows the possibility of producing entangled states between the 
Cooper box and any other dynamical system, possibly macroscopic, 
to which it can be coupled. Examples include coupling the Cooper box 
to another large superconducting island \cite{marquardt}, a superconducting 
resonator\cite{buisson} and to a micromechanical 
gate electrode \cite{armour}, which could take the form of a 
cantilever or bridge-like structure. In the present work, we examine the 
effect of a mechanical gate electrode on the energy levels of a Cooper 
box. In particular, we consider what might be viewed as a mechanical analogue 
of the Lamb shift, in which the quantum zero-point motion of the 
mechanical oscillator modifies the level separation between the Cooper 
box states.

The Hamiltonian for the Cooper box-coupled mechanical gate electrode 
system is
\[H=4 E_{\rm C} [n_{\rm g}-(n+1/2)]\hat{\sigma}_{z} -\frac{1}{2} 
E_{\rm J}\hat{\sigma}_{x} +\hbar\omega\hat{a}^{\dagger}\hat{a} 
-4E_{\rm C}n_{\rm g}^{\rm m}\frac{\Delta x_{\rm zp}}{d} 
(\hat{a}+\hat{a}^{\dagger})\hat{\sigma}_{z},
\]
where $E_{\rm C}=e^{2}/2C_{\rm J}$ is the single-electron charging 
energy,   $n_{\rm g}=-(C_{\rm g}^{\rm c} V_{\rm g}^{\rm c}+
C_{\rm g}^{\rm m}V_{\rm g}^{\rm m})/2e$ 
is the dimensionless, total gate charge with control gate voltage  
$V_{\rm g}^{\rm c}$ and 
mechanical gate electrode voltage $V_{\rm g}^{\rm m}$  
chosen such 
that $n_{\rm g}$ is close to $n+1/2$ for some $n$ (so that only Cooper charge 
states $\left|n\right.\rangle \equiv \left(_0^1 \right)$ and
$\left|n+1\right.\rangle \equiv \left(_1^0 \right)$ play a role), 
$n_{\rm g}^{\rm m}=-C_{\rm g}^{\rm m}V_{\rm g}^{\rm m}/2e$, $E_{\rm J}$ 
is the Josephson coupling energy, 
$\Delta x_{\rm zp}$ is the zero-point displacement uncertainty of the 
mechanical gate electrode, $d$ is the mechanical electrode-island 
gap, and $\omega$ is the frequency of the fundamental 
flexural mode of the mechanical electrode. We
assume $C_{\rm J}\gg C_{\rm g}$ and $d\gg\Delta x_{\rm zp}$. 

Neglecting the coupling between the Cooper box and mechanical 
electrode, the energy eigenvalues are $E^{(0)}_{0,N}=-\Delta 
E(\eta)/2 +N\hbar\omega$ and $E^{(0)}_{1,N}=+\Delta 
E(\eta)/2 +N\hbar\omega$, where $\Delta  E(\eta)=E_{\rm 
J}/\sin\eta$ with the mixing angle $\eta =\tan^{-1}(E_{\rm 
J}/[8E_{\rm C}(n+1/2-n_{\rm g})])$. To second order in the coupling, 
we have 
\begin{eqnarray*}
    E^{(2)}_{1,N}-E^{(2)}_{0,M}&=&\Delta 
E(\eta)\left[1+32(N+M+1)\left(\frac{\Delta x_{\rm 
zp}}{d}\right)^{2}\right.\\
&\times&\left.\left(\frac{E_{\rm C}^{2}}{(\Delta 
E(\eta))^{2}-(\hbar\omega)^{2}}\right)\sin^{2}\eta\ {n_{\rm 
g}^{\rm m}}^{2}\right]+(N-M)\hbar\omega.
\end{eqnarray*}
Notice that we require $\eta\neq 0$ in order for the coupling to 
modify the energy levels. If $E_{\rm J}=0$, then the only effect of 
the coupling of the Cooper box to the mechanical oscillator is a 
shift of the harmonic potential to the left or to the right, 
depending on the state of the Cooper box. With $E_{\rm 
J}\neq 0$ and $\hbar\omega < \Delta E(\eta)$, we see that the 
interaction with the mechanical oscillator increases the gap 
between the Cooper box levels.  

Let us now estimate the magnitude of the possible gap increase under realisable 
conditions, supposing the mechanical oscillator to be in its ground  
state  and assuming also that $\hbar\omega \ll \Delta E(\eta)$. 
Josephson energies for Cooper boxes are typically in the tens of 
micro-eV range, 
translating to tens of GHz which exceeds by at 
least an order of magnitude the fundamental frequencies of realisable 
micromechanical oscillators. For the oscillator undergoing zero-point 
motion ($N=M=0$), 
the gap increase is then approximately $32 {n_{\rm g}^{\rm m}}^{2} (\Delta x_{\rm 
zp}/d)^{2} (E_{\rm C}/E_{\rm J})^{2}\sin^{4}\eta$. Considering, for 
example, $E_{\rm C}=100~\mu{\rm eV}$, $E_{\rm J}=10~\mu{\rm eV}$, 
$n+1/2-n_{\rm g}=0.01$, $\Delta x_{\rm zp}=10^{-2}$~{\AA}, 
$d=0.1~\mu{\rm m}$, and $n_{\rm g}^{\rm m}=10$, the gap increase is about 
$10^{-5}$, likely too small to be detected.  

If, on the other hand, the mechanical oscillator is in a thermal state, then 
for the same parameter values the gap increase is approximately 
$10^{-5} (2\bar{N} +1)$, where $\bar{N}$ is the 
thermal-averaged occupation number.  Considering, for example, a 
fundamental frequency $\nu=50~{\rm kHz}$ and temperature $T=30~{\rm 
mK}$, we have $\bar{N}\approx 2.5 \times 10^{4}$ giving a significant gap 
increase of about $0.25$. 

A possible way to probe the effect of the mechanical oscillator 
thermal motion on the Cooper box levels is measure the photon-assisted 
Josephson quasiparticle (PAJQP) tunneling current  dependence on total gate 
charge $n_{\rm g}$ \cite{nakamura2}. As the mechanical gate voltage
$V_{\rm g}^{\rm m}$ is turned on, increasing $n_{\rm g}^{\rm m}$, 
we would expect the PAJQP current 
peak to the left (right) of the main Josephson quasiparticle (JQP) 
tunneling current peak to shift towards the right (left), signalling the 
increasing gap between the $n$ and $n+1$ Cooper box levels. At the 
same time, the PAJQP peaks should broaden due to the thermal motion 
of the mechanical oscillator.

A proper analysis of the quasiparticle tunneling current is required which 
includes the corrections to the Cooper box energy levels due to the 
coupling to the mechanical oscillator. A suitable starting point is 
the analysis of the experiment of Nakamura {\it et al.} 
\cite{nakamura} given in Ref.\ \cite{choi}. This will be the subject 
of a future investigation.

This work was supported in part  by the NSA
and ARDA under ARO contract number DAAG190110696, and by 
the EPSRC under Grant No. GR/M42909/01.

\end{document}